
\input phyzzx

\nopubblock
\voffset=24pt
\advance\vsize by -.5cm
\advance\hsize by -1.5cm
\titlestyle{THE ONE DIMENSIONAL MATRIX MODEL AND STRING THEORY \foot{Based
on lectures delivered at the Spring School on Superstrings at ICTP,
Trieste, April, 1992.}}

\author{Sumit R.Das}

\address{Tata Institute of Fundamental Research, \break
Homi Bhabha Road, Bombay 400005}

\nopagenumbers

\centerline{\bf ABSTRACT}

{\narrower\narrower\smallskip\singlespace
We discuss the basic features of the double scaling limit of the
one dimensional matrix model and its interpretation as a two dimensional
string theory. Using the collective field theory formulation of the model
we  show how the fluctuations of the collective field can be interpreted
as the massless "tachyon" of the two dimensional string in a linear
dilaton background. We outline the basic physical properties of the theory
and discuss the nature of the S-matrix. Finally we show that the theory
admits of another interpretation in which a certain integral transform of
the collective field behaves as the massless "tachyon" in the two
dimensional string with a blackhole background. We show that both the
classical background and the fluctuations are non-singular at the black
hole singularity.
\smallskip}

\vskip .2in

\def\cZ{{\cal Z}}
\def\half{{1 \over 2}}
\def\paa{\partial}
\def\onen{{1 \over N}}
\def\Pieta{\Pi_{\eta}}
\def\paq{\partial_q}

\def\ps{\psi (x,t)}
\def\psd{\psi^{+}(x,t)}
\def\half{{1 \over 2}}
\def\pat{\partial_t}

\def\pax{\partial_x}

\def\inx{\int dx}

\def\Pp{P_{+}}
\def\Pm{P_{-}}
\def\Ppm{P_{\pm}}

\def\sqtwo{{\sqrt{2}}}

\def\ba{{\bar A}}

\def\cZ{{\cal Z}}

\def\stwo{{\sqrt{2}}}

\def\ps{\psi}
\def\psd{\psi^{\dagger}}
\def\pat{\partial_t}
\def\pax{\partial_x}

\def\half{{1 \over 2}}

\def\sin{{\rm sin}}
\def\cos{{\rm cos}}
\def\sinh{{\rm sinh}}
\def\cosh{{\rm cosh}}

\def\sqbeta{{\sqrt{\beta}}}
\def\quarter{{1 \over 4}}
\def\llog{{\rm log}}
\def\gst{g_{st}}
\def\tT{{\tilde T}}
\def\inx{\int d^2x}
\def\pax{\partial_x}
\def\half{{1 \over 2}}
\def\pat{\partial_t}
\def\tphi{{\tilde \phi}}

\def\smu{{\sqrt{\mu}}}

\def\twa{{2 \over \alpha '}}
\def\pau{\partial_u}
\def\pav{\partial_v}

\def\Win{W_{1+\infty}}

\def\stwo{{\sqrt{2}}}

\def\tom{T_\omega}

\def\rega{{\rm Region~I}}
\def\regb{{\rm Region~II}}

\Ref\MM{V. Kazakov, Phys. Lett 150B (1985) 282;
V. Kazakov, I. Kostov and A. Migdal, Phys. Lett. 157B (1985) 282; F.
David, Nucl. Phys. B257 (1985) 45; J. Ambjorn, B. Durhuus and J. Frohlich,
Nucl. Phys. B257 (1985) 433.}

\Ref\THOFT{G. 't Hooft, Nucl. Phys. B72 (1974) 461.}

\Ref\BIPZ{E. Brezin, C.
Itzykson, G. Parisi and J. Zuber, Comm. Math. Phys.59 (1978) 35.}

\Ref\DSCAL{M. Douglas and S. Shenker, Nucl. Phys.
B335 (1990) 635; D. Gross and A. Migdal, Phys. Rev. Lett. 64 (1990) 717;
E. Brezin and V. Kazakov, Phys. Lett. 236B (1990) 144.}

\Ref\KMIG{V. Kazakov and A.A. Migdal, Nucl. Phys. B311 (1988) 171.}

\Ref\DEQONE{D. Gross and N. Milkovic, Phys. Lett. B238 (1990) 217; E. Brezin,
V. Kazakov and Al. B. Zamolodchikov, Nucl. Phys. B338 (1990) 673; P.
Ginsparg and J. Zinn-Justin, Phys. Lett. 240B (1990) 333; G. Parisi, Phys.
Lett. 238B (1990) 209.}

\Ref\GKK{D. Gross and I. Klebanov, Nucl. Phys. B344 (1990) 475.}

\Ref\DJEV{S.R. Das and A. Jevicki, Mod. Phys.
Lett A5 (1990) 1639}

\Ref\JEVSAK{A. Jevicki and B. Sakita, Nucl. Phys. B165 (1980) 511.}

\Ref\SWAD{A. Sengupta and S.R. Wadia, Int. J. Mod. Phys. A6
(1991) 1961.; D. Gross and I. Klebanov, Nucl. Phys. B352 (1990) 671.}

\Ref\DNW{S.R. Das, S. Naik and S.R. Wadia, Mod. Phys. Lett. A4 (1989) 1033.}

\Ref\DDW{S.R. Das, A. Dhar and S.R. Wadia, Mod. Phys. Lett. A5 (1990) 799.}

\Ref\OTHER{J. Polchinski, Nucl. Phys. B331 (1989) 123 ;
F. David and E. Guitter, Europhys. Lett. 3 (1987) 1169 ;
A. Dhar, T. Jayaraman, K.S. Narain and S.R. Wadia,
Mod. Phys. Lett. A5 (1990) 863.}

\Ref\SHAP{J. Shapiro, Nucl. Phys. B184 (1981) 218.}

\Ref\KOST{I. Kostov, Phys. Lett. 215B (1988) 499.}

\Ref\VIJAY{L.C.R. Wijewardhana, Phys. Rev. D25 (1982) 583.}

\Ref\AB{I. Andric and V. Bardek, Phys.
Rev. D32 (1985) 1025.}

\Ref\POLB{J. Polchinski, Nucl. Phys. B346 (1990) 253.}

\Ref\KARS{D. Karabali and B. Sakita, CUNY Preprint
(1991).}

\Ref\POLYDIS{A. Polyakov, Mod. Phys. Lett. A6 (1991) 635.}

\Ref\MSW{G. Mandal, A. Sengupta nd S. Wadia, Mod. Phys. Lett. A6 (1991) 1465.}

\Ref\DJR{K. Demeterfi, A. Jevicki and J.P. Rodrigues, Nucl. Phys. B362
(1991) 173; B365 (1991) 199.; Mod. Phys. Lett. A35 (1991) 3199.}

\Ref\MORE{ G. Moore, Nucl. Phys. B368 (1992) 557.}

\Ref\DIFR{P. Di Francesco and D. Kutasov, Phys. Lett. B261 (1991) 385.}

\Ref\GKSM{D. Gross and I. Klebanov, Nucl. Phys. B359 (1991) 3.}

\Ref\POLSM{J. Polchinski, Nucl. Phys. B 362 (1991) 125.}

\Ref\DMW{A. Dhar, G. Mandal and S.R. Wadia, TIFR Preprints TIFR-TH/91-61
and TIFR /TH/ 92-40; Talks by A. Dhar and S. Wadia in this meeting.}

\Ref\MPR{G. Moore, R. Plesser and S. Ramgoolam, Yale Preprint YCTP-P35-91
(1991); G. Moore and R. Plesser, Yale Preprint YCTP-P7-92 (1992);
A. Jevicki, J. Rodrigues and A.J. van Tonder, Brown Preprint BROWN-HET
-874 (1992).}

\Ref\AJEV{J. Avan and A. Jevicki, Phys. Lett B266 (1991) 35; Phys. Lett.
B272 (1990) 17; Mod. Phys. Lett. A7 (1992) 357.}

\Ref\DDMW{S.R. Das, A. Dhar, G. Mandal and S.R. Wadia, Int. J. Mod. Phys.
A7 (1992); Mod. Phys. Lett. A7 (1992) 71; Mod. Phys. Lett. A7 (1992) 937.}

\Ref\MORSEIB{G. Moore and N. Seiberg, Int. J. Mod. Phys. A7 (1992) 2601.}

\Ref\POLWIN{D. Minic, J. Polchinski and Z. Yang, Nucl. Phys. B369 (1992) 324}

\Ref\WGROUND{E. Witten, Nucl. Phys. B373 (1992) 187; I. Klebanov and
A. Polyakov, Mod. Phys. Lett. A6 (1991) 3273}

\Ref\DAS{S.R. Das, TIFR Preprint TIFR-TH/92-62.}

\Ref\BLACKA{G. Mandal, A. Sengupta and S.R. Wadia, Mod. Phys. Lett. 6
(1991) 1685}

\Ref\BLACKB{E. Witten, Phys. Rev. D44 (1991) 314.}

\Ref\MART{E. Martinec and S. Shatashvili, Nucl. Phys. B368 (1992) 338.}

\Ref\EGUCHI{ T. Eguchi, H. Kanno and S.K. Yang, Newton Institute Preprint
NT-92004 (1992).}

\Ref\BATEMAN{ {\it Tables of Integral Transforms}, Volume II (Bateman
Manuscript Project) p 145 (McGraw-Hill, 1954).}

\Ref\DVV{R. Dijkgraaf, E. Verlinde and H. Verlinde, Nucl. Phys. B371
(1992) 269. }

\Ref\DMWBH{A. Dhar, G. Mandal and S.R. Wadia, TIFR-TH/92-63.}

\Ref\ELL{J. Ellis, N.E. Mavromatos and D. Nanopoulos, Phys. Lett. B272
(1991) 261. ;
J. Ellis, N.E. Mavromatos and D. Nanopoulos, Phys. Lett. B267
(1991) 465; R. Brustein and S. de Alwis, Phys. Lett. B272 (1991) 285.}

\chapter{\bf Matrix Models and Random surfaces}

One of the most fruitful approaches to the study of random surfaces and
noncritical strings is to
consider a dynamical triangulation of a two dimensional surface and then
searching for continuum limits.

Discretized random surfaces may be viewed
as Feynman diagrams of field theories of matrix fields, or Matrix models
[\MM]. This really follows from 't Hooft's discovery that the large-N
expansion of certain matrix models ia a {\bf topological} expansion
[\THOFT].

Consider a matrix field theory of a $N \times N$ hermitian matrix
$M_{ij}(X^\mu)$ in $d$ dimensions. The action is taken to be of the form
$$S_M = N [\int d^dx \int d^dy~{\rm Tr} \half M(x) G^{-1}(x,y) M(y) + \int
d^dx~{\rm Tr} V(M)] \eqn\fiftysevena$$
Here $V(M)$ is some polynomial potential which is of the form
$$ V(M) = \sum_k g_k ~M^k \eqn\fiftyeighta$$
In \fiftysevena\ we have an yet unspecified propagator for the
matrix, $G(x,y)$. The above field theory has a global $U(N)$ symmetry $M
\rightarrow UMU^{-1}$ for an arbitary constant $U(N)$ matrix $U$. We shall
be interested in computing quantities which are invariant under this
symmetry.

We shall study this model in the large-$N$ expansion, i.e. we shall expand
all quantities in inverse powers of $N$, considering all the couplings
$g_k$ to be of order unity. This means that to every order of ${1 \over
N}$ we have to sum an infinite number of feynman diagrams which have all
possible powers of the couplings, but each of which is of the same order
of $\onen$.

One can represent each propagator
of the matrix field by a {\bf double line}, each line corresponding to an
index of the matrix $M_{ij}$ (hence called an index line). The two index
lines have arrows attached opposite direction [\THOFT].
In terms of double lines a typical feynman diagram looks like a two
dimensional
surface which is broken up into tiles. Each tile is bounded by a closed
index loop and adjacent tiles are attached to each other along the
boundaries. This, of course, is a discretization of a surface with the
following correspondence
$$\eqalign{ &{\rm faces} \rightarrow {\rm index ~loops} \cr &
{\rm edges} \rightarrow {\rm propagators} \cr &
{\rm vertices} \rightarrow {\rm interaction ~vertices}}$$
Vacuum diagrams of the field theory represent {\it closed} two dimensional
surfaces. For general correlators we will have surfaces with points whose
positions are not integrated over.
Since $N$ appears as a multiplicative factor in front of the action, each
propagator brings a factor of $\onen$, each index loop brings a factor of
$N$ since indices have to be summed over for invariant qauntities, while
each vertex also brings a factor of $N$. Thus the total $N$-dependence in
any diagram is given by
$$ (N)^{V+T-L} = N^{\chi} $$
where $V$ denotes the number of vertices, $T$ the number of faces and $L$
the number of edges. $\chi$ is the Euler characteristic.
This shows that the $\onen$ expansion is a {\it topological
expansion}, i.e. for a given order of $\onen$ one has to sum all possible
diagrams of a given topology.
This, of course, is a feature of string perturbation theory and $\onen$
as a string coupling.

Let us examine a bit more detail the expression for the free energy (just
as an example) for our matrix field theory. For simplicity we shall
consider an action which has a monomial interaction. For example we could
take only the coupling $g_4$ defined above to be non-zero.
A given feynman diagram is evaluated in the standard way by joining
together vertices with propagators. The real space expression is
$$ g_4^V \int \prod_i dx_i~\prod_{<ij>} G(x_i,x_j) $$
where $V$ is the number of vertices.
The product is over all pairs of points connected by a propagator.
The result for the free energy is
obtained by a sum over all feynman diagrams and finally a sum over all
orders of $\onen$
$$ F_{MM} = \sum_{\chi}(\onen)^{\chi}
\sum_{diagrams(N)}g_4^V \int \prod_i dx_i~\prod_{<ij>}
G(x_i,x_j) \eqn\fiftynine $$
The sum over feynman diagrams in \fiftynine\ is split in accordance with
the $\onen$ expansion,i.e. first we sum over all diagrams in a given order
of $N$ and then sum over the various orders.
Note that since each feynman diagram is a discretization of a two
dimensional surface, the sum over all closed feynman diagrams is precisely
thes sum over all possible discretizations of a surface. The expression
\fiftynine\ is precisely of the same form as the {\it partition function}
of a set of scalar fields $x^\mu$  on a dynamically discretized random
surface. Since the number of vertices is a measure of the area, the
cosmological constant of the random surface theory is given by
$$ \mu = - {\rm log}~ g_4 $$
In fact if we choose the propagator to be of the exponential form
$$ G(x_i,x_j) = e^{-(x_i - x_j)^2} \eqn\sixty$$
we have precisely the discretized version of the Polyakov string. This is
because if one is able to define a continuum limit for this model, the
factor $\prod_{<ij>} G(x_i,x_j)$ becomes $exp~[-\int d^2 \xi~(\partial
x)^2]$ while the sum over all feynman diagrams becomes a functional
integral over the worldsheet metric. Finally the sum over $\chi$ is the
genus expansion so the {\bf string coupling constant} $g_{st} = {1 \over
N}$.
It is worth remembering that
the equivalence is between the {\it free energy} of the matrix field
theory and the {\it partition function} of the random surface model.
The relationship between the cosmological constant and the couplings of
the matrix model shows that the continuum limit of the random surface
theory corresponds to singularities of the free energy of the matrix model.
\vskip 1.0cm

\centerline{\underbar{The Double Scaling Limit}}

The main point about matrix models is that they provide a {\it
non-perturbative} definition of random surface theories. As argued above
the $\onen$ expansion of matrix models gives the genus expansion of the
random surface theory, i.e. the string perturbation expansion. However the
matrix model exists independently of its large-N expansion.
This gives a hope that one
could study non-perturbative effects in string theories using matrix
model methods.

In fact the continuum treatment of David, Distler and Kawai already
indicates what should be done in order to define a continuum limit which
does not proceed genus by genus. Recall that for a conformal field theory
with central charge $c$ coupled to two dimensional gravity, the
partition function behaves as $(\mu^{\alpha})^{\chi}$, where $\alpha$ is
determined in terms of $c$ and positive for $c \leq 1$ and $\chi$ is the
euler characteristic. In
the matrix model we have seen that $\mu$ in the above expression
corresponds to $(g - g_c)$. Furthermore, the string coupling constant
$g_{st}$ comes raised to the power $-\chi$. Thus near the critical point
the full partition function as a sum over genus has a generic form
$$ \cZ \sim \sum_{\chi} [{(g-g_c)^{\alpha} \over g_{st}}]^{\chi}~Z_{\chi}
\eqn\sixthree$$
The $\onen$ corrections to the planar limit of several solvable matrix
models also led to an identical form for the free energy.

The equation \sixthree\ shows that
it is possible to define a {\bf double scaling limit}
$$ g \rightarrow g_c~~~~~~~g_{st} \rightarrow 0 ~~~~~~
{(g-g_c)^{-\alpha} \over g_{st}} = {\rm fixed} \eqn\dscal$$
In the matrix model $g_{st} = \onen$ so that this double scaling limit is
described by both $g \rightarrow g_c$ as well as $N \rightarrow \infty$.
In this limit {\it all} orders of genus expansion contribute equally and
one clearly has a {\bf non-perturbative definition} of the continuum limit
of the random surface theory.

The crucial idea of double scaling limit was discovered in
[\DSCAL]. In these papers the one matrix model, whose genus zero solution
was already known from the classic paper of [\BIPZ],
was solved {\it directly} in this limit. The remarkable feature
of their work is that the specific heat of the model satisfies a
non-linear integrable differential equation (different for different
multicritical points) and the flow between various multicritical points is
described by euqtions of the KdV hierarchy. We shall not desribe this
seminal work. Rather we shall directly proceed to describe the double
scaling limit of the matrix field theory in {\it one} dimension.

\chapter{\bf The $d = 1$ Matrix model : Double scaling.}

We start with the matrix field theory in one dimension which we will
identify as time. In other words we are dealing with some matrix quantum
mechanics. As seen above the matrix model
action which corresponds to the Polyakov action for strings is of the form
$$ S = \int dt {\rm Tr}[M(t)~e^{\paa_t^2}~M(t) + V(M(t))] \eqn\sixfour$$
For the model to have a smooth large-$N$ limit the couplings in the
expansion of the polynomial potential $V(M)$ have to be specific, viz
$$ V(M) = \sum_{k=2} {g_k \over N^{k-2 \over 2}}~M^k \eqn\jone$$
By rescaling $M$ we can always choose $g_2 = 1$ and we will also rename
$g_4 \equiv g$ and define $\beta = {N \over g}$.
Field theories with exponential propagators
are cumbersome to work with.
In one dimension, however, there is a great simplification. This is
because in one dimension the behavior of loop integrals do not depend on
the behavior of the propagator at small distances. This means that we can
keep only the lowest terms in the expansion of $e^{\paa_t^2}$. This brings
the action to a conventional form with the standard kinetic term for
matrix quantum mechanics.
$$ S = \int dt {\rm Tr}[(\paa_tM)^2 + V(M)] \eqn\sixfive$$
In the lowest order of the large-N expansion this model was solved in
[\BIPZ]. Its interpretation in terms of random surfaces is discussed for
example in [\KMIG].
We shall discuss the double scaling limit of this problem by considering
first a potential of the form
$$V(M) = M^2 + {g \over N} M^4 \eqn\sixsix$$
Very soon we shall see that only some generic properties of the potential
are important in the double scaling limit. First let us rescale the matrix
$M$ such that all $N$ dependence appears as an overall factor in the
action.
$$S = \beta \int dt {\rm Tr}[(\paa_tM)^2 + v(M)] \eqn\sixseven$$
where $v(M)$ denotes the potential $V(M)$ with $g$ being set to be equal
to $1$.  Since we have a
standard kinetic term it is convenient to consider the corresponding
Hamiltonian. This is given by
$$
\hat{H} = \rm{Tr} \left(- {1 \over 2 \beta}~ {\partial^{2} \over
\partial M^{2}} + \beta v(M)\right)\eqn\sixeight$$
We shall restrict ourselves to the sector of the theory which is singlet
under the symmetry group of $U(N)$ rotations. One can then reformulate the
problem in terms of the eigenvalues
$\lambda_i$ of the matrix $M$.
In
the change of variables from the matrices $M$ to the eigenvalues one has a
jacobian which is the Van der Monde determinant [\BIPZ].
$$ \prod_{ij} dM_{ij} = \prod_i d\lambda_i~ (\prod_{i \ne j} (\lambda_i -
\lambda_j))^2 \eqn\sixten$$
and one can redefine the wave functions to absorb the determinant
$\prod_{i \ne j} (\lambda_i -\lambda_j)$. The new wave functions are then
slater determinants, i.e. fermionic many body wave functions. The
hamiltonian acting on these fermionic wavefunctions reads
$$ \hat{H} = \sum_i [{1 \over 2 \beta}~ {\partial^{2} \over
\partial \lambda_i^{2}} + \beta v(\lambda_i)] \eqn\sixnine$$
All we have are $N$ fermions living in an external potential $v(\lambda)$.
It is convenient to talk of a reduced hamiltonian $h$
given by
$$ \hat{H} = \beta h  = \beta \sum_i h_i$$
Let $e$ denote the energy levels and $\rho(e)$ the density of states of
the reduced single particle hamiltonian $h_i$.
Since we have $N$ fermions the fermi level $\mu_F$ is determined by
$$ \int^{\mu_F}_{0} de~\rho (e) = N \eqn\seventya$$
While the ground state energy of the system is given by
$$ E_{gs} =\beta \int^{\mu_F}_{0} de~e~\rho (e) \eqn\seventy$$
Note that this is the energy of the original system, not the value of the
reduced hamiltonian, which explains the factor of $\beta$ in \seventy. The
equations \seventy\ and \seventya\ determine the ground state energy of
of the model.

Let us first consider the semiclassical limit.
It is clear from the form of the hamiltonian $\hat{H}$ that the inverse of
$\beta$ acts as a Planck's constant. Thus the semiclassical limit of the
model is given by $\beta \rightarrow \infty$ at fixed value of the
coupling $g$. This is, of course the large-$N$ limit at fixed coupling, or
the {\it planar limit}.
The form of the reduced potential $v(\lambda)$ relevant for our
case is sketched below
\vskip 5.0cm
The crucial feature of the potential is the existence of maxima at $\pm
\lambda_c$. In the lowest order of the semiclassical approximation one can
use the Bohr-Sommerfeld quantization rule to determine the energy levels
of the hamiltonian. Thus if $e_n$ denotes the $n$-th energy level of the
{\it reduced hamiltonian} $h$ one has
$$ \int_{-\lambda_+}^{\lambda_+} d\lambda~{\sqrt{2(e_n - v(\lambda)}}
= (n - \half){1 \over \beta} \eqn\jtwo$$
The factor of ${1 \over \beta}$ appears on the right hand side because
this is the Planck's constant for our problem. The limits of integration
over $\lambda$ are as usual the endpoints of the classical orbits. The
density of states then follows:
$$ \rho(e) = \beta \int {d \lambda \over {\sqrt{2(e - v(\lambda)}}}
\eqn\seventhree $$
Criticality
is obtained when the fermi energy $\mu_F$ approaches $v(\lambda_c)=\mu_{Fc}$,
which happens when the cosmological constant $g$ approaches a critical value
$g_c$.
A crucial role is played by the quantity
$$\mu = \mu_{Fc} - \mu_{F} \eqn\sevenone$$
Near criticality it is now straightforward to find the relationship
between $\mu$ and the quantity $(g - g_c)$.
Then the filling condition \seventya\ implies
$$ {dg \over d \mu_F} = \int_{-\lambda_0}^{\lambda_0}
{d \lambda \over {\sqrt{2(\mu_F - v(\lambda)}}} \eqn\sevenfour$$
Now in the limit where $\mu_F \rightarrow \mu_{Fc}$, the classical
particle spends most of its time near the turning point and the integral
in \sevenfour\ receives most of its contribution from the turning point.
It is then easy to evaluate the integral by expanding the potential around
the point $\lambda_c$ and show that
$$ {dg \over d \mu_F} \sim {\rm log}~\mu \eqn\sevenfive$$
near the critical limit. Integrating \sevenfive\ one concludes that
$$ \Delta g = (g - g_c) \sim \mu {\rm log}~\mu \eqn\sevensix$$
The value of the ground state energy also follows easily:
$$E_{gs}^{0} -{1 \over 4 \pi} (\beta \mu)^2 ~{\rm log}~\mu \eqn\jthree$$
The expression for the
density of states \seventhree\ shows why the limit $\mu \rightarrow 0$
corresponds to a continuum limit. This is because it is fairly easy to
check that the density of states at the fermi level $\rho(\mu_F)$ diverges
as ${\rm log}~\mu$ in the critical limit (this is in fact the content of
\sevenfive). Thus the excitations above the ground state, which are
particle hole pairs because of the fixed number of fermions constraint,
will have a continuous spectrum of energies - a signature of continuum
physics.

It is also possible to the next correction to the ground state energy. We
will not detail the procedure here, but simply quote the answer
$$ E_{gs}^{(1)} = {1 \over 12 \pi}~{\rm log}~\mu \eqn\jfour$$
The expressions for the ground state energy in the first two orders of the
large-$N$ expansion suggests what the double scaling limit could be. In
the double scaling limit the energies in every order should be comparable.
Comparing \jthree\ and \jfour\ we see that this can be {\it almost}
achieved if we consider the limit
$$ \beta \rightarrow \infty,~~~~~\mu \rightarrow 0,~~~~~ (\beta \mu) =
{\rm fixed}\eqn\seventwo$$
The problem is the additional ${\rm log}~\mu$ piece in \jfour. We shall
see that this logarithmic factor contains important physics. For the
moment we ignore its presence and declare the double scaling limit of the
model to be defined by \seventwo.

It is remarkable that in the double scaling limit the problem simplifies
considerably. To see this consider solving the single particle Schrodinger
equation with the hamiltonian $h$
for energies near the fermi level. We have to solve
$$ [{1 \over 2 \beta^2}~ {\partial^{2} \over
\partial \lambda^{2}} + v(\lambda)] \psi (\lambda) = \mu_F \psi (\lambda)
\eqn\sevenseven$$
We now expand the potential around the critical point $\lambda_c$ and
perform a further rescaling by defining the variable $x$
$$ x = {\sqrt{\beta}}(\lambda_c - \lambda) \eqn\seveneight$$
Then the above Schrodinger equation for a wavefunction at the fermi level
reads
$$ {1 \over \beta}[{1 \over 2 }~ {\partial^{2} \over \partial x^2} + \beta
\mu + \half v''(\lambda_c) x^2 - {1 \over 6 {\sqrt{\beta}}} v'''(\lambda_c)
x^3 + \cdots ] \psi (x) = 0 \eqn\sevennine$$
Since $\beta \mu$ is fixed in the double scaling limit, it is clear that
the terms in the potential involving powers of $x$ higher than $2$ are
suppressed by inverse powers of ${1 \over {\sqrt{\beta}}}$. Furthermore,
we shall be interested in the case of the potential having a maximum, i.e.
$v"(\lambda_c) < 0$.
In the double scaling limit, therefore, one has a
problem of fermions living in an {\it inverted} harmonic oscillator
potential
\foot{In the above considerations we have assumed that $v''(\lambda_c) \ne 0$.
This is certainly the situation for generic potentials with a critical
point. However, for more special potentials one could have
${d^k v \over d \lambda^k} = 0$ for all $k < m$. The various scaling
behaviors of the density of states, the ground state energy etc. are now
dependent on $m$, leading to different exponents \Ref\DDSWB{S.R. Das, A.
Dhar, A. Sengupta and S.R. Wadia, Mod. Phys. Lett. A5 (1990) 891.}.
However, the physics of these models are not understood very well. In what
follows we shall exclusively with the generic case of $m = 2$.}.

The resulting quantum mechanical problem thus has a hamiltonian
$$H_{ds} = -\half {\partial^2 \over \partial x^2} + \beta v(\lambda_c)
-\half x^2 \eqn\jfive$$
where we have rescaled fields to have $v''(\lambda_c) = -1$. The height of
the potential is of order $\beta$ at the maximum, and it is easily seen
that in terms of these scaled coordiantes $x$ the zero of the original
potential, which is at $\lambda = 0$ is of order ${\sqrt{\beta}}$ from the
origin $x=0$. In the following we shall measure the energies from the top
of the hump of the potential, which means that the term $\beta
v(\lambda_c)$ is not present in the hamiltonian \jfive.

As it stands the problem is not well defined since one has a potential
which is bottomless. To define the problem one has to put a wall to bound
the potential. Now, in the critical limit a classical particle moving in
an orbit with energy equal to the fermi energy spends most of its time
near the turning point. In the double scaling limit this region is blown
up, which is why only the $x^2$ term is important. The higher order terms
in the potential are suppressed by powers of ${1 \over {\sqrt{\beta}}}$
and become important only when $x \sim O(\sqbeta)$. Thus one may put a
wall at $\vert x \vert = \Lambda \sim \sqbeta$ without interfering with
the universal physics.

The density of states of the problem may be obtained in a number of ways.
We will give the treatment of Brezin et. al. in [\DEQONE]. We are
interested in solving the Schrodinger equation for energies not far from
the top of the potential hump. If $\xi$ denotes the energy measured from
the top we are interested in $1 << \xi << \Lambda^2$. Thus there is a
large coordinate range ${\sqrt{\xi}} << x << \Lambda$ where one is (a) far
from the turning point so that semiclassical approximation for the wave
function is valid and (b) far from the wall so that it is sufficient to
retain the quadratic term in the potential. The wavefunctions of the
inverted harmonic oscillator potential are parabolic cylinder functions
which has the asymptotics
$$ \psi (x) \sim {1 \over {\sqrt{x}}}~\sin~[\quarter x^2 + \xi~
\llog(x) + \half \phi(\xi)] \eqn\jsix$$
where
$$ \phi (\xi) = -\half i~\llog~{\Gamma(\half + i \xi ) \over \Gamma (\half
- i\xi)} + O(e^{-2\pi\xi}) \eqn\jseven$$
On the other hand far from the turning point, i.e. for $x >> {\sqrt{\xi}}$
one can use the WKB wave function
$$ \psi (x) \sim {1 \over {\sqrt{x}}}~\sin~[\quarter x^2 + \xi~
\llog({x \over \Lambda}) + \half \Phi({\xi \over E_0}, g_i)] \eqn\jeight$$
where the function $\Phi$ depends on the details of the potential (the
nature of the wall, hence on the values of the coefficients of the higher
terms in the Taylor expansion of the potential, $g_i$). $E_0$ is the
height of the potential. Since $\xi << E_0$ in our region of interest we
can set ${\xi \over E_0} = 0$ so that the entire information of the
details of the potential far from the hump is encoded in a single
parameter $\Phi (0,g_i) \equiv \Phi_0$
independent of the energy level. Matching the two
wave functions \jsix\ and \jeight\ one has the quantization condition
$$ \Phi_0 - \xi~\llog~\Lambda - \phi (\xi) = \pi~n $$
where $n$ is an integer. The density of states follow immediately
$$ \rho = -{1 \over \pi}({\partial \phi \over \partial \xi} +
\llog~\Lambda)$$
where $\phi(\xi)$ is given in \jseven. The density of states may be now
expanded in inverse powers of $\xi$ and used in the expressions \seventy\
and \seventya\ to obtain the following expansion for the ground state
energy [\DEQONE]
$$E_{gs} = -{1 \over 4\pi} (\beta \mu)^2 {\rm ln} \mu +
{1 \over 12\pi}{\rm ln} \mu + \sum_{m=1}^{\infty} C_m (\beta \mu )^{-m}
\eqn\eighty$$
where $C_m$ are some numerical coefficients.
We also write down the expression for the finite temperature free energy $F$
for the singlet sector at temperature $T$ [\GKK]
$$ {F \over T} = -{1 \over 4\pi T} (\beta \mu)^2 {\rm ln} \mu +
{1 \over 12}{\rm ln} \mu (\pi T + {1 \over \pi T}) + .... \eqn\eightyone$$
The higher terms are all finite in the double scaling limit and respect the
duality symmetry $ \pi T \rightarrow {1 \over \pi T}$.
Note that unlike the situation in the single matrix model, the expressions
for $E_{gs}$ and $F$ are not finite in the double scaling limit. Rather
the two leading contributions have the well known "logarithmic scaling
violations". One of our main aims is to understand these scaling violations.

\chapter{\bf Collective Field Theory}

The spacetime interpretation of noncritical string theory hinges on the
fact that the liouville mode acts as an extra dimension in target space
[\DNW, \DDW, \OTHER]. Since the one dimensional matrix model becomes,
in the critical
limit, the Polyakov string with one worldsheet scalar (the target space
time) one expects that the true target space dimension must be two. At
first sight this appears confusing. In the matrix model representation the
integration over the liouville mode is really the sum over all feynman
diagrams and some extra dimension does not seem to appear explicitly.
Nevertheless the formulation of the theory as a model of $N$ mutually
noninteracting fermions in an external potential immediately shows that
there is an extra dimension - viz. the space of eignvalues which we have
been calling $\lambda$. In fact the double scaled model is by definition
equivalent to a $1+1$ dimensional field theory of non-relativistic
fermions. One might wonder whether the eigenvalue direction $\lambda$ or
its scaled version $x$ is the additional dimension in target space. We
shall now argue that this is not quite true - the "liouville mode" is
related to $x$, but not $x$ itself.

Though the model is naturally written in terms of fermionic fields, the
excitations are all bosonic. This is because $N$ is fixed so that the
excitations are all bosonic particle-hole pairs. In this section we will
show that application of the Collective Field Theory approach leads to a
direct physical interpretation of the one dimnesional matrix model
[\DJEV].

The main idea is to rewrite the problem in terms of
invariant variables
using the method of collective fields developed in [\JEVSAK].
Correlation functions of any
number singlet operators may be written in terms of the collective fields,
$$\phi (x) = \int {dk \over {2\pi}} e^{-ikx} {\rm Tr}~(e^{ikM}) = \sum_i
\delta (x - \lambda_i)\eqn\eightytwo$$
$\phi(x)$ is simply the density of eigenvalues $\lambda_i$.

We now make a change of variables from the fields $M_{ij}$ to the
collective fields $\phi (x)$.
The field $\phi(x)$ is, of course,
constrained to satisfy:
$$  \int dx ~ \phi(x) = N \eqn\eightythree $$
The change of variables is made using the procedure of [\JEVSAK]. We
also perform the rescalings  $x \rightarrow \sqrt{\beta} x, \phi \rightarrow
\sqrt{\beta} \enskip \phi$.
One finally gets a Hamiltonian [\DJEV]
$$
H_{\phi} =  \int dx  ( {1 \over {2 \beta^2}}
\partial_{x} \pi(x) \phi(x) \partial_{x} \pi(x)
{}~ + \beta^2 V_0 + \Delta V) \eqn\eightfour$$
where $\pi(x)$ is the momentum conjugate to the field $\phi(x)$, and
$$
V_{0} = {\pi^{2}\over 6} \phi^{3} + (v(x) - \mu_{F}) \phi(x)\eqn\eightfive$$
$$
\Delta V = {1\over 2} \int_{y=x}dx \phi(x) \partial_{x} \partial_{y} ln
\vert x-y \vert \eqn\eightsix$$
In this hamiltonian, the constraint \eightythree which has a $g$ instead of
$N$ on the right hand side, has been implemented by a lagrange multiplier
$\mu_F$.
The lagrangian which follows from the above hamiltonian may be written as
$$  {\cal L} = (\beta)^2 \int dx
\biggl\{ {1 \over 2} \partial_{x}^{-1} \dot{\phi}{1\over{\phi}}
\partial_{x}^{-1} \dot{\phi}      - V_{0} (\phi (x,t)) - \Delta V
\biggr\}\eqn\eightseven$$

\noindent \underbar{Leading Order}

It is clear that ${1 \over\beta}$ is the bare string coupling constant. In the
leading order of the WKB expansion the free energy is dominated by the
saddle point solution for $\phi(x)$ which is given by
$$
\phi_{0}(x) = {1\over{\pi}} \sqrt{2(\mu_{F} - v(x))}\eqn\eighteight$$
while its integral must be equal to $g$.
In the critical limit of small $\mu$
one may easily show that the energy evaluated at the saddle point
coincides with the leading order ground state energy in \eighty.

\noindent \underbar{The Liouville mode and the Spectrum}

The lagrangian density \eightseven\ is clearly the lagrangian of a two
dimensional field theory with $x$ as the extra dimension.
However $x$ is not quite the liouville mode we are looking for. To find
the liouville mode we have to study the fluctuations around the saddle
point solution [\SHAP]
$$ \phi(x) = \phi_0 (x) + {1 \over \beta}
{\partial \eta \over \partial x}
\eqn\eightnine$$
It is convenient to introduce the variable $q$ defined by:
$$
q = {1 \over{\pi}}\int^{x} {dx\over{\phi_{0}(x)}}, \quad\quad  dq =
{dx\over{\phi_{0}}}\eqn\ninety$$
Note that $q$ is simply the time of flight of a classical particle moving
in the potential $v(x)$. In the following we will use the specific form of
the potential
$$
v(x) = x^2 - 2x^4 \eqn\nineone$$
The results are of course independent of the detailed form of the
potential provided the maximum is generic, i.e. the second derivative
$v''(x_0)$ is non-vanishing.

The range of $q$ is given by $ -L < q < L$ where
$4L$ is the time period of classical motion. In the critical limit $L
\rightarrow \infty$ as $L = -{1 \over 4} {\rm ln} \mu$ (as follows from
the saddle point solution above). The action quadratic in the fluctuations
$$
S_{2} = \pi^{3} \int dt \int_{-L}^{L} dq \biggl[ {1\over 2}
({\partial_{t}\eta})^2 - {1\over 2} (\partial_{q}
\eta)^{2}\biggr]\eqn\ninetwo$$
which is the lagrangian of a massless scalar field in two dimensions.

That is exactly what is expected from the continuum theory for $d=1$! The
one dimensional non-critical string may be considered as a two dimensional
critical string theory with a linear dilaton background in the liouville
direction [\DNW, \DDW].
The equations of motion for the modes are then obtained in the
standard manner by considering the corresponding backgrounds in the
worldsheet theory and requiring Weyl invariance. In two dimensions, the
scalar is the only full-fledged field. Consider a tachyon background
$T(t,\phi)$. Its equation of motion which follows from the weyl invariance
condition is
$$B(T) = (-\partial_\phi^2 + 2 \sqtwo \partial_\phi + \partial_t^2 + 2)
T(t,\phi) + V'[T(t,\phi)] = 0 \eqn\fiftya$$
The potential $V(T)$ is nonuniversal but starts with a term cubic in $T$.
In our case one has a cosmological constant background as well. This means
that there is a time-independent tachyon background. The precise nature of
this background depends on the non-universal nonlinear pieces in \fiftya.
However, as argued in [\POLB] the features of the background for $\phi
\rightarrow \pm \infty$
can be nevertheless obtained.
$$\eqalign{&T_0 (\phi - \phi_0) \rightarrow 1 - a~e^{(2 -
\sqtwo)\phi}~~~~~~~~ \phi \rightarrow - \infty \cr &
T_0 (\phi - \phi_0) \rightarrow b~\phi~e^{-\sqtwo\phi}~~~~~~~~
\phi \rightarrow + \infty}\eqn\pone$$
Here $\phi_0$ is the center of the kink like configuration. The energy of
this configuration is infinite, so one needs to put an upper cutoff to
$\phi$, which may be taken to be $\phi = 0$. Thus all $\phi < 0$. The
value of the field at $\phi =0$ is taken to be $T(0,t) = \Delta$, where
$\Delta$ will be identified with the {\it bare} cosmological constant on
the world sheet. The aim is now to minimize the energy maintaining the
boundary condition at $\phi = 0$. The resultant minimum energy
configuration is given by ${\bar T}(\phi,t) \equiv T(\phi - {\bar \phi})$
where ${\bar \phi}$ is defined by
$$ - b~ {\bar \phi}~e^{-\sqtwo~{\bar \phi}} = \Delta \eqn\ptwo$$
Note that as $\Delta \rightarrow 0^+$, ${\bar \phi} \rightarrow - \infty$.
Comparison with the relationship between the cosmological constant and the
fermi energy in the matrix model immediately leads to the identification
${\bar \phi} = {1 \over \sqtwo} \llog~\mu $. To find the equation
satisfied by the excitation one has to expand around this ground state
solution ${\bar T}$. However, for small $\Delta$, the classical
configuration is practically zero for all ${\bar \phi} < \phi < 0$ so the
linear equation for the fluctuation ${\tilde T} \equiv T - {\bar T}$ is
given once again by the linear part of the equation \fiftya. One can now
make a field redefinition ${\tilde T} \rightarrow e^{-\sqtwo \phi} {\tilde
T}$ so that the linearized equation for the new ${\tilde T}$ is
$$ [-\partial_\phi^2 + \pat^2] {\tilde T} = 0 \eqn\cvone$$
which represents a massless scalar field in two dimensions. However, very
close to the "wall" ${\bar \phi}$ there is an additional term in the
linearized equation of motion \cvone\ which is ${\bar T}~{\tilde T}$. This
additional term is proportional to the cowsmological constant $\Delta$.

Comparison of the linearized equation of motion of the collective field
fluctuation $\eta$ and the tachyon fluctuation ${\tilde T}$ far away from
the wall suggestes the interpretation that the time of flight variable $q$
is the liouville mode and $\eta$ the tachyon field. However the fact that
the tachyon fluctuations in the linear dilaton- cosmological constant
background has departures from the massless free field behaviour {\it even
at the linearized level} indicates that this interpretation can be only
approximate. A better interpretation has been suggested in [\MORSEIB].
Consider the macroscopic loop operator which reads, in the language of
collective field theory
$$ W(p,t) = \int_{\smu}^{\infty} dx~e^{-px}~\phi (x,t) \eqn\cvtwo$$
The parameter $p$ may be regarded as the invariant length of a large loop
on the worldsheet. It may be easily checked that the classical value of
$W(p,t)$, obtained by substituing $\phi_0$ for $\phi (x,t)$ in \cvtwo\ is
given by ${\smu \over p}~K_1 (\smu p)$ where $K_\nu (z)$ stands for the
standard modified Bessel function. The fluctations of $W$ around this
classical value may be seeen to obey the equation
$$ [(p \partial_p)^2 - \pat^2 - \mu~p^2]~{\tilde W}(p,t) = 0 \eqn\cvthree$$
This equation seems to indicate that $\llog~p$ is the quantity that has to
interpreted as the liouville mode and $W(p,t)$ as the tachyon field. This
is because \cvthree\ has the right qualitative features of the equation
for the tachyon fluctuation around a linear dilaton - cosmological
constant background. Far away from the "wall" this reduces to the standard
massless scalar equation, with $\llog~p \sim \phi$ and close to the wall
the departure is proportional to the cosmological constant. One may wonder
whether the theory can be written in terms of the $W(p,t)$. In principle
it can be done, but the result is nonlocal and cumbersome to work with and
for obtaining the consequences of the theory it is much easier to work
with the fluctautions of the collective field $\eta$ itself.

The boundary conditions on the fluctuation $\eta(q,t)$ are determined from
the time independence of the constraint, i.e. ${d \over dt} (\int dx
\phi (x)) = 0$, which
leads to Dirichlet boundary conditions on $\eta$ : $ \eta(-L,t) = \eta(L,t)
= 0$.  The eigenfunctions are therefore
$$\eta_n(q) = {1\over \sqrt L} \sin ({n\pi q\over L})~~{\rm or}~~~
= {1\over \sqrt L} \cos (n+{1\over 2}) {\pi q\over L}\eqn\ninethree$$
where $n=0,1,2,\cdots$. The frequencies are
$$\omega_{j} = {j \pi \over 2L} = j\omega_{c}; \qquad j = 0,1,2....
\eqn\ninefour$$
\noindent The propagator is then
$$
D(t-t^{\prime}; q, q^{\prime}) = \int {dE \over {\pi}} e^{iE(t-t^{\prime})}
D(E,q,q')
$$
where
$$
D(E,q,q') =\sum_{j} \quad {\eta_{j}(q)
\eta_{j}(q^{\prime})\over E^{2} - \omega_{j}^{2} +
i \epsilon}.\eqn\ninefive$$
\noindent In the scaling limit we have ~$L \rightarrow \infty$ and we define
continuum momenta
$p =  {n\pi\over{2L}}$. The propagator now becomes the standard massless
scalar propagator in two dimensions.
The dispersion relation becomes $E^{2} - p^{2} = 0$.

Using the above basic two point function it is straightforward to evaluate
the class of two-point functions in the matrix model
$$
< Tr~M^p(t)  Tr~M^q(0) >_{c} =      \int dx \int dx' x^p~x^q~~
\partial_x \partial_{x'}< \eta (x,t) \eta (x',t) > \eqn\ninesix$$
Using \ninethree\ and \ninefive\ it is easy to check that
the result is identical to that obtained from a direct calculation of the
above correlator in the matrix model [\KOST].

\noindent \underbar{One loop free energies}

To obtain the one-loop (torus) free energy at zero temperature we need to
calculate the expression
$$
\Delta E_0 = \pi^3 \int dq [\partial_q \partial_{q'} G(q,q')]_{q=q'}
\eqn\nineseven$$
where $G(q,q')$ is the standard propagator following from the action
\ninetwo\
$$
G(q,q') = -{1 \over 2\pi^3} {\rm ln} \vert q - q' \vert \eqn\nineeight$$
One must also add the term $\Delta V$ in the original
collective field lagrangian. To treat the various singular
pieces, we make a change of variables in \nineeight\ to the original
variable $x$. Using the definition of $q(x)$ one finds that the
singularity as $x \rightarrow y$ cancels between $\Delta V$ and
\nineeight,
leaving with the finite answer
$$
\Delta E_{1} = {1\over{24\pi}} \int  {v^{\prime\prime}(x)dx\over
{[2(\mu_{F}  - v(x))]^{1\over 2}}} \eqn\ninenine$$
Using the form of the potential one can find the singular (as
$\mu \rightarrow 0$) piece in the one loop free energy density to be
$$\Delta E_1 = {1 \over 12 \pi} {\rm ln} \mu  \eqn\hundred$$
which agrees with the corresponding term in \eighty. The expression
\hundred\ also agrees with the leading WKB correction obtained in
[\VIJAY], and the
collective field calculation in [AB].

To obtain the free energy at finite temperature one has to add to
$\Delta E_1$ the contribution of the thermal free energy of a massless
scalar field in (1 + 1) dimensions having a massless dispersion relation.
At temperature $T$ the free energy is easily seen to be
$$ F/T = {\pi LT \over{3}} = - {\pi T \over{12}} \ln \mu \eqn\hone$$
The total thermal free energy is obtained by adding \hundred\ and \hone\
$$ {F \over T} = {1 \over 12} \ln  \mu (\pi T + {1 \over \pi T})$$
which displays duality and agrees with \eightyone.

Thus the genus zero and one free energies are proportional to ${\rm ln}
\mu$ simply because the effective field theory lives in a box of length $L
\sim {\rm ln} \mu$.

\noindent \underbar{Interactions}

To obtain the full structure of the interactions in the theory it is best
to work in the hamiltonian framework. We thus work with the collective
hamiltonian and perform perturbations around the classical
solution. It is convenient to introduce a field $\xi$ and its canonically
conjugate momentum $\Pi_\xi$ and rescale $x$
$$ x \rightarrow (\half \mu)^\half x~~~~\phi = (\half \mu)^{-\half}\pax
\xi~~~~\pax \Pi_\phi = - (\half \mu)^{-{3 \over 2}} \Pi_{\xi} \eqn\fone$$
and introduce
$$ \Ppm = -g_{st}^2 \Pi_{\xi} (x,t) \pm \pi \pax \xi (x,t) \eqn\hsix$$
The hamiltonian \eightfour\ then acquires a simple form
$$
H =   { 1 \over 2 \pi g_{st}^2} \int dx~[{1 \over 6} (\Pp^3 - \Pm^3) -
\half x^2 (\Pp - \Pm) + \mu (\Pp - \Pm) ] \eqn\hseven$$
In \hseven\ we have imposed the constraint \eightythree\ by means of a
lagrange multiplier $\mu$.
The fluctuations are then carried out by writing
$$
\Ppm = \pm \phi_0 + {g_{st} \over \phi_0} \eta_{\pm} \eqn\height$$
where
$$\eta_{\pm} = - \Pi_{\eta} \pm \partial_{\tau} \eta \eqn\hnine$$
The fluctuation hamiltonian is then
$$H =   {1 \over 2\pi} \int dq \lbrace [(\eta_{+})^2 + (\eta_{-})^2] +
{g_{st} \over 3 \phi_0^2} [(\eta_{+})^3 - (\eta_{-})^3] \rbrace \eqn\hten$$

The discusssion of the one loop free energy indicates that normal ordering
of this hamiltonian removes the singularities already present in H. The
normal ordered form of the Hamiltonian turns out to be
$$ H =    \int dq ~[\half(:\Pieta^2 + (\paq \eta)^2:) +
{1 \over 2\beta \phi_0^2}
(:\Pieta (\paq \eta) \Pieta + {1 \over 3} (\paq \eta)^3:  + V']
\eqn\heleven$$
where the extra finite term
$$V' = - {1 \over 12} [{\phi_0 '' \over \phi_0} -  \half ({\phi_0' \over
\phi_0})^2 ] (\paq \eta) + {1 \over 24} [ {(\phi_0 ')^2 \over \phi_0} - 2
\phi_0 ''] \eqn\htwelve$$
replaces the infinite term $\Delta V$. The interactions in the hamiltonian
are purely cubic and there is a finite tadpole. The finite field
independent term in \htwelve\ is of course the one loop ground state
energy.

This exact form of the hamiltonian has been derived from the fermionic
field theory of the matrix model [\SWAD].
The remarkable feature of
the hamiltonian is that the interactions are strong only at the boundary.
For our potential $\phi_0 (q)$ may be obtained in terms of standard
elliptic functions, and in the critical limit, one can show that
$${1 \over \phi_0^2} \sim {\rm cosh}^4(q)\eqn\hthirteen$$
The coupling has the exponential dependence on the liouville direction as
has been found in the continuum theory [\DNW, \POLB] .
The coupling
grows near the boundaries.  In fact in the double scaling limit the
coupling at the boundary is held fixed
$$g_{st} = {1 \over \beta} \exp( 4 L) = {1 \over \beta \mu} = {\rm fixed}.
\eqn\hfourteen$$
and equal to the string coupling constant $\gst$
This
immediately provides a qualitative explanation of why there are no ${\rm
ln} \mu$ factors in front of the higher genus contributions to the free
energy. The higher genus contributions are purely due to interactions
which fall off exponentially away from the boundary;
therefore no overall volume factor is present.

In several recent papers Demeterfi, Jevicki and Rodrigues [\DJR]
has developed a systematic hamiltonian perturabtion theory based on
\heleven\ and \htwelve\ and have calculated the two loop ground
state energy and the two loop finite temperature free energy. The results
are {\it completely finite} and
in exact agreement with the results of the matrix model. However the
calculations seem to depend on the type of regularization used. In
[\MSW] the two point function of the eigenvalue density was calculated
to one loop using a different regularization and the answer was found to
be divergent. This indicates the presence of counterterms which are both
finite and infinite in this regularization scheme. The bosonization
technique employed in [\KARS]  also seems to indicate the presence of
higher terms in the hamiltonian. In [\DMW] the fermionic theory has been
bosonized in terms of the quantum distrtibution function in phase space
and it has been concluded that the collective field theory results only if
some approximations are made.
However, the fact that the prescription
for handling infinities used in [\DJR] leads to the correct expression
of the thermodynamic free energy at two loops indicates that the cubic
hamiltonian is sufficient to calculate such quantities, at least in string
perturbation theory. It is, however, possible that the non-perturbative
extension defined by the collective field theory may be defined by the
non-perturbative extension defined by the fermionic field theory.

\noindent\underbar{The Tree Level S- Matrix}

The tree level S-Matrix of the theory has been obtained from many different
approaches. The most direct approach is to use the standard perturbation
theory techniques in the collective field theory [\MSW, \DJR] or directly
use the fermionic field theory [\MORE]. In this section we shall summarize
the approach of [\POLSM]. The main motivation to explain this approach is
that it uses the fermi fluid picture and provides a useful intuition about
the scattering process, and in formulating an exact bosonization of the
system [\DMW].

Our model is a model of $N$ fermions which do not interact with one
another living in an external inverted harmonic oscillator potential. The
second quantized form of the action is therefore
$$ S = {1 \over \gst}
\int dt~dx~\psd(x,t)[i\pat + \half \pax^2 + \half x^2]\ps(x,t)
\eqn\zone$$
The collective field theory action has an overall factor of ${1 \over
g_{st}^2}$ whereas the fermionic action has an overall factor of ${1 \over
g_{st}}$ (recall that $g_{st} = {1 \over \beta \mu}$). This implies that
the classical limit of the collective field theory corresponds to the
classical {\it single particle} limit in the fermionic theory, viz. to the
classical motion of particles in the external inverted harmonic oscillator
potential.

The ground state of the many-fermion system corresponds to the filled
fermi sea - this corresponds to the classical ground state of the
collective field theory.  In the two dimensional phase space $(p,x)$
of the single particle the fermi surface is a hyperbola $\half(p^2 - x^2)
= \beta \mu$. An excited state means an excitation of a
particle hole pair. This would correspond to a deformation of the
hyperbola. In the collective field theory this represents a general field
configuration with energy higher than the ground state. Thus a general
state of the bosonic field is in one to one correspondence with a profile
of the fermi surface.

To look for classical solutions of the collective field theory one has to,
therefore, look for deformations of the fermi surface.
Let $\sigma$ be a parameter which denotes a point on the fermi surface.
The motion of the point in phase space may be obtained directly from the
Hamiltonian evolution equations:
$$ x(\sigma,t) = - a(\sigma)~\cosh~(t - b(\sigma))~~~~p(\sigma,t) = -
a(\sigma)~\sinh~ (t - b(\sigma)) \eqn\ktwo$$
We can use the freedom of reparametrization of the fermi surface to set
$b(\sigma) = \sigma$ so that the general solution is
$$ x(\sigma,t) = - a(\sigma)~\cosh~(t - \sigma)~~~~p(\sigma,t) = -
a(\sigma)~\sinh~ (t - \sigma) \eqn\kthree$$
There is a direct translation of the quantities in the single particle
phase space and the collective field theory. The density of eigenvalues of
the matrix is the collective field $\phi (x,t) = {\pax \xi \over g_{st}}$.
At the classical level the density of points in phase space $u(x,p,t)$ is
given by (in our normalization)
$$ u(x,p,t) = {1 \over 2\pi g_{st}} \theta (-\beta \mu - \half(p^2 - x^2))
\eqn\kfour$$
(Recall that we are measuring energies from the top of the hump). This
means that the density of fermions should be
$ \psi^{\dagger} \psi = \int_{p_- (x,t)}^{p_+ (x,t)} {dp \over 2 \pi \gst}
= {p_+ (x,t) - p_(x,t)- \over 2 \pi \gst}$
where $p_\pm$ are the upper (lower) edge of the fermi surface for a given
value of $x$.  The momentum density in the collective field theory is simply
$\Pi_\xi \pax \xi (x,t)$, which is given by
$\int_{p_- (x,t)}^{p_+ (x,t)} {dp~p \over 2 \pi \gst^2}
= - {p_+^2 (x,t) - p_-^2 (x,t) \over 4 \pi \gst^2} $. Comparison with
\hsix\ immediately shows that we should
identify $p_\pm$ with $P_\pm$. In fact integration of the single particle
hamiltonian over phase space and use of this identification immediately
leads to the classical collective field hamiltonian.

It is now clear that the general solution to the collective field theory
may be obtained directly from \kthree. In fact \kthree\ is a parametrized
form of the fermi surface which may be alternatively written in terms of
$p_\pm (x,t)$ and hence $P_\pm (x,t)$.

To obtain the tree level $S$-matrix we restrict our attention to the left
portion of the potential hump, i.e. $x < 0$. For the inverted harmonic
oscillator potential the
time of flight variable $q$ is
$$ q = - \llog~(-x) \eqn\kseven$$
Consider now a point on the deformed fermi surface with some value of
$\sigma = \sigma_0$. At some time $t_1$ this corresponds to some value of
$x (\sigma_0,t_1) = x_0$ given by the equation \kthree\ and denotes a particle
moving with the momentum $p_0$ given in \kthree\ at the position $x_0$ at time
$t_1$. We will take the time $t_1$ such that this corresponds to a
particle moving towards the potential hump, so that $x_0 < 0, p_0 > 0$. Then
eliminating $p$ from \kthree\ one has
$$ x_0 - {\sqrt{x_0^2 - a^2}} = -a~e^{t_1 - \sigma_0} \eqn\keight$$
We will be interested in the large $\vert x \vert$ region, i.e. $q
\rightarrow - \infty$. In this region one can solve for $\sigma_0$ in terms
of $q_0$ and $t_1$ from \keight\ to obtain
$$ \sigma_0 = t_1 - q_0 - \half \llog~({1 \over 4} a^2) \eqn\knine$$
The particle will travel to the potential and bounce back. Eventually at
some later time $t_2$ it will again cross the point $x_0$, or equivalently
$q_0$ with a momentum $- p_0$ since it is now travelling away from the
potential. From \kthree\ this means
$$ t_2 - \sigma_0 = - (t_1 - \sigma_0) \eqn\kten$$
Using \knine\ we thus have the time difference between successive
crossings of the same spatial point as
$$ t_2 - t_1 = - 2q_0  - \llog~({1 \over 4} a^2) \eqn\keleven$$

Now in the asymptotic region $x \rightarrow - \infty$ one has $p_+
\rightarrow \vert x \vert$ and $p_- \rightarrow - \vert x \vert$. It is
therefore natural to define new objects $\epsilon_\pm$ by
$$ p_\pm = \pm e^{-q} \mp e^q~\epsilon_\pm \eqn\ktwelve$$
It is easy to check that
$$ \epsilon_+ = (p_+ + x)~x \eqn\ktwelve$$
Using the parametrized solutions \kthree, the solution for $\sigma$ in
the asymptotic region and \ktwelve\ one gets
$$ a^2 = 2(p_+ + x)~x = 2 \epsilon_+ (t_1 - q_0) \eqn\kthirteen$$
In the process we are considering, i.e. a particle coming in and bouncing
back, with momentum $p_0$ at the position $x_0$ while going in
and a momentum $-p_0$ at the same position while going back we have
$$ p_0 = e^{-q_0} - e^{q_0} \epsilon_+ (q_0,t_1) ~~~~~~~~
-p_0 = -e^{-q_0} + e^{q_0} \epsilon_- (q_0,t_2) \eqn\kfourteen$$
which means
$$ \epsilon_+ (t_1,q_0) = \epsilon_- (t_2,q_0) \eqn\kfifteen$$
The edges of the fermi sea, $p_\pm (x,t)$ satisfy the equation
$$ \pat p_\pm = x - p_\pm \pax p_\pm \eqn\kfifteen$$
{}From the definition of the $\epsilon_\pm$ it may be easily checked that in
the asymptotic region $q \rightarrow -\infty$ they satisfy the equation
$$ [\pat^2 - \partial_q^2]~\epsilon_\pm = 0 \eqn\ksixteen$$
These are purely right or left moving waves. Clearly for the incoming
particles, one has $\epsilon_+ (t - q)$ whereas for the bounced wave one has
$\epsilon_- (t + q)$. Thus we have from \kfifteen, using \keleven\ and
\kthirteen\
$$\epsilon(t_1 - q_0) = \epsilon_- [t_1 - q_0 + \llog~2 - \llog~\epsilon_+
(t_1 - q_0)] \eqn\kseventeen$$
Since this equation is valid for all values of $t_1$ and $q_0$ we might as
well remove the subscripts on these quantities.

To see what the equation \kseventeen\ means in terms of the scattering of
the fundamental excitations of the theory we have to recall that the
quantities $p_\pm$ are nothing but the chiral combinations of momenta and
fields of the collective field theory, $P_\pm (x,t)$. We thus express the
quantities $\epsilon_\pm$ in terms of the field $S(q,t)$ and its
canonically conjugate momentum $\Pi_S (q,t)$ as
$${1 \over {\sqrt{\pi}} \gst}\epsilon_\pm (q,t) = \pm \Pi_S(q,t) -
\partial_q S(q,t) \eqn\keighteen$$
Note that $S(q,t)$ is {\it not} the same as the fluctuation of the
collective field $\eta (q,t)$ used earlier. In fact the vacuum values are
$\epsilon_\pm = \half$ which corresponds to $S(q,t) = -{q \over 2
{\sqrt{\pi}}~\gst}$. We thus make the mode decomposition of the field
$S(q,t)$ as
$$S(q,t) =  -{q \over 2 {\sqrt{\pi}}~\gst} + i\int_{-\infty}^{+\infty}
{dk \over 2^{{3 \over 2}} \pi k} [ \alpha_+ (k)~e^{ik(t - q)} +
\alpha_- (k)~e^{ik(t + q)}] \eqn\knineteen$$
The $\alpha (k)$ are annahilation operators for right moving modes, while
$\alpha_- (k)$ are annahilation operators for the left moving
modes.

The relation \kseventeen\ thus provides a relation between the left and
right moving modes. The crucial point is that the equation \kseventeen\ is
a non-linear equation so that the relation between the left and right
moving modes is nonlinear as well. This is the key to the question as to
why there could be non-trivial scattering of the bosons while the fermions
from which they come from are free. To find out the relation between the
$\alpha(k)$ and $\alpha_-(k)$ one has to expand $\epsilon_\pm$
around their classical value $\half$
$$\epsilon_\pm (t \mp q) = \half + {\tilde{\epsilon}}_\pm (t \mp q \mp
\llog~2) $$
Then \kseventeen\ may be used to solve for ${\tilde{\epsilon}}_+$ as a
power series in ${\tilde{\epsilon}}_-$, which in turn leads to the
following relation between $\alpha_\pm (k)$
$$\eqalign{\alpha_+ (k) = & \alpha_- (k) - ik {\sqrt{2 \pi}} \gst
\int_{-\infty}^{+\infty} {d k_1 \over 2 \pi} \alpha_-(k_1)~\alpha_-(k -
k_1) + \cr & (ik - k^2) {4 \pi \over 3} \gst^2 \int \int {dk_1~dk_2 \over
4 \pi^2} \alpha_-(k_1)~\alpha_-(k_2)~\alpha_-(k - k_1 - k_2) + \cdots}
\eqn\ktwenty$$

It is now easy to read out the S-matrix from \ktwenty.
The main point is that since we are working on a half-line, incoming modes
are right moving while outgoing modes are left moving. For example, when
we have two particles with momenta $k_1$ and $k_2$ coming in and two
particles with momenta $k_3$ and $k_4$ going out one has to simply
evaluate the quantity $<0|\alpha_- (k_3)~\alpha_- (k_4)~
\alpha_+^\dagger (k_1)~\alpha_+^\dagger (k_2)|0>$ by using \ktwenty. The
result for the $T$ matrix is
$$T = - {i g_{st}^2 \over 16 \pi}   {\sqrt{k_1k_2k_1'k_2'}} (|k_1 + k_2| +
     |k_1 - k_1'| + |k_1 - k_2'| - 4i) \eqn\hfifteen$$
The same S-matrix has been obtained in [\MSW - \MORE]. More significantly the
same S-matrix has been reproduced from a continuum calcualtion in [\DIFR].

Note the absence of any momentum conserving delta function in the result.
This is because the interactions break translation invariance. It has been
suggested by Polyakov [\POLYDIS] that the general form of the S-Matrix in
general non-critical string theories is
$$S = R(k,k') + { A(k,k') \over (\sum k_i - \sum k'_i)} \eqn\hsixteen$$
where $k$ and $k'$ denote incoming and outgoing momenta. If the momenta
are conserved then the second term in \hsixteen\ is proportional to
the volume of the spatial direction. In the $d=1$ theory $A(k,k')$ is
identically zero ! In fact this follows that the two chiralities are
decoupled in the collective field hamiltonian [\GKSM]. For some further
understanding of the S- Matrix see [\MPR]

\chapter{\bf The $W_{\infty}$ Symmetry}

The model we have been discussing has a remarkable set of global
symmetries which form an infinite dimensional algebra [\AJEV - \POLWIN].
The symmetries are best described in the fermionic formulation since it is
only in this formulation the {\it exact} symmetries are known [\DDMW].

Since we have a system of $N$ fermions in an external potential, the
problem is completely integrable and has an infinite number of conserved
and mutually commuting charges [\SWAD] with their conserved currents.
It turns out that these charges are the Cartan subalgebra of the full
algebra of symmetry generators forming $\Win$

Consider the fermionic action \zone. It is enlightening to regard the
argument $x$ as an index for the fermionic field so that we have
$$ S = \int dt \sum_{x,y}\psd_x(t)[i\pat \delta(x-y) + \ba_{xy}(t)
]\ps_y(t) \eqn\xone$$
where the matrix $\ba_{xy}(t) = \half(\pax^2 + x^2)~\delta(x-y)$. This
immediately shows that the double scaled matrix model may be viewed as a
theory of fermions in the fundamental representation of $U(\infty)$ living
in an {\it external} gauge field $\ba_{xy}(t)$.
In fact we can consider any general, but fixed, background.
Thus the free energy of
the system would be unchanged if we consider a different background gauge
field related to the original one by the gauge transformation
$$\ba \rightarrow  {\cal U} (t) \ba {\cal U}^{\dagger} (t) + i {\cal U}
(t) \pat {\cal U}^{\dagger} (t) \eqn\xthree$$
where we have suppressed indices. In \xthree\ $\ba$ is a hermitian matrix
in the $(x,y)$ space and ${\cal U}$ is a unitary matrix. For arbitrary
${\cal U}$ this is {\bf not} a symmetry of the action (for some given $\ba$)
since the gauge field is not a dynamical variable. However, there may be
some special ${\cal U}$'s for which the gauge field $\ba$ does not change
: these are then symmetries of the fermionic action. To identify these
symmetries consider an infinitesimal transformation ${\cal U} =
e^{\epsilon}$ where $\epsilon$ is again a matrix. Then the transformation
of the gauge field is $\ba \rightarrow \ba + {\rm D}_t \epsilon$ where
${\rm D}$ is the covariant derivative in the given background. Thus the
symmetries of the theory are given by solutions of
$$ {\rm D}_t~\epsilon (t) = 0 \eqn\wone$$
For our double scaled potential a set of solutions are
$$ \epsilon_{xy}(t) = e^{-(r-s)t}~[(x - i\pax)^r, (x +
i\pax)^s]_{+}~\delta (x-y) \eqn\wtwo$$
where $r,s$ are positive integers or zero.
The corresponding charges which generate these symmetries on the fermionic
fields are
$$ W_{rs} = e^{-(r-s)t}
\int dx~\psd (x,t)~[(x - i\pax)^r, (x + i\pax)^s]_+~\ps (x,t) \eqn\wthree$$
These charges form a closed algebra which is isomorphic to the full $\Win$
algebra. In the {\it classical} limit, i.e. for small $\gst$ this reduces
to the algebra of area preserving diffeomorphisms on the plane. It is this
algebra which has been found in the continuum formulation [\WGROUND].
Note that for $r \neq s$ the charges are time dependent. This means that
these charges do not commute with the hamiltonian, though the action is
kept invariant. The commuting set of infinite charges $W_{rr}$ are
those found in [\SWAD].
The
implications of the symmetry and the algebra will be discussed in detail
in the lectures of A. Dhar in these proceedings.

We end this section by writing down the symmetries in the collective field
formulation. At the classical level, they may be easily obtained by using
the correspondence between the single particle operators and collective
field theory operators discussed in the previous section. The charges are,
therefore [\AJEV]
$$ \omega^{(r,s)} =  \inx \int_{\Pm}^{\Pp} dp~ (p+x)^r (p-x)^s~e^{-(r-s)t}
\eqn\bostwo$$
It is straightforward to check that the collective field hamiltonian \hseven\
is given by $\omega^{(1,1)}$ while the Poisson bracket algebra
$$ \lbrace \omega^{(r,s)}~,~\omega^{(r',s')} \rbrace_{{\rm PB}} = -(r's-sr')
\omega^{(r+r'-1,s+s'-1)} \eqn\bosthree$$
which is the $w_{1+\infty}$ algebra.
In a similar way it is straightforward to check that
${d \omega^{(r,s)} \over dt} = 0$.

In the quantum theory, one has to define a suitable ordering for the
charges defined above. One such definition is given in
[\AJEV] who conclude that the algebra is unchanged in the quantum
theory. This is in contradiction to what we found in the fermionic
formulation. It is probable that a different ordering is required which
reproduces the correct algebra.

\chapter{\bf The Matrix Model as a Black Hole background}

In the above sections we have presented ample evidence that the
double scaling limit of the matrix
model represents a linear dilaton - cosmological constant background of
the two dimensional critical string theory.

In this section we will argue
that it is possible to interpret the matrix model rather differently : as
a black hole background of the string theory [\DAS]. More precisely,
we show that a certain {\it linear}
integral transform of the fluctuations of the
collective field obeys the same linearized
classical equation of motion as that of a massless scalar
in the black hole background of the two dimensional critical string.
Our connection is related to the observation in [\MART] that the coset
model of [\BLACKB] may be recast as a liouville theory by going over to
the space of field momenta on the worldsheet, but rather different from
pervious attempts to describe the black hole background by modifying the
collective field theory [\ELL].
To facilitate comparison with standard conventions of the black hole
metric we shall rescale the fermi level $\mu \rightarrow 2 \mu$ so that
the classical solution to the collective field is
$\phi_0 (x) = {\stwo \over \pi} (x^2 - 4\mu)^{\half}$.

It may be easily checked that the fluctuation ${\tilde
\phi}$ then satisfies the following equation
$$ \half \pat^2 \tphi = \tphi + 3x \pax \tphi + (x^2 - 4 \mu) \pax^2 \tphi
\eqn\qfive $$
Now consider the following transform of the collective field
$$ T(u,v) = \int_0^\infty dp \int_{-\infty}^\infty dt~
e^{ip[e^t v + e^{-t} u]}~
\int_{2 \smu}^{\infty} dx~ e^{-px} \phi (x,t)
\eqn\qeleven$$
In other words we are considering a transform of the macroscopic loop
operator introduced in \cvtwo.
In an obvious notation we will call the transform of $\phi_0$ $T_0(u,v)$,
while the transform of the fluctuation will be referred to as ${\tilde T}
(u,v) $. Using the Dirichlet boundary conditions on the field $\eta$
introduced earlier ($\tphi = \pax \eta$) it may be checked that ${\tilde
T}$ satisfies the following equation
$$ [4(uv + \mu) \pau \pav + 2(u \pau + v \pav) + 1] {\tilde T}(u,v) = 0
\eqn\qtwelve $$
This is precisely the equation of the massless tachyon moving in a black
hole background of two dimensional critical string theory written in
Kruskal like coordinates. The invariant form of the equation is
$$ \nabla^2 T - 2 \nabla T \cdot \nabla D + (\twa) T = 0 \eqn\qthirteen$$
where $\nabla$ denotes target space covariant derivative and $D$ is the
dilaton background. $\alpha '$ is the string tension. In Kruskal like
coordinates the black hole solution in the small $\alpha '$ limit
has the metric and dilaton fields
[\BLACKA]
$$ \eqalign{G_{uv} = & G_{vu} = {1 \over 2(\twa~uv + a)}~~~G_{uu} = G_{vv}
= 0 \cr & D(u,v) = -\half~{\rm log}~(\twa~uv + a)}\eqn\qfourteen$$
The parameter $a$ is the mass of the black hole.
Substituting \qfourteen\ in \qthirteen\ and comparing the resulting equation
with  \qtwelve\ we get for the black hole mass
$$ a = \twa~\mu \eqn\qfifteen$$
The same relation holds in the connection between $SL(2,R)$ coset model
and liouville theory proposed in [\MART].

The field $T(u,v)$ is defined in the entire two dimensional plane. We will
now show explicitly that $T(u,v)$ thus defined gives the correct solution
of \qtwelve\ in all the regions of the $(u,v)$ plane. To do this it is
convenient to define coordinates $(r, \theta)$ in the four regions as
follows
$$ \eqalign{ & u = r~e^\theta~~~~v = r~e^{-\theta}~~~~{\rm for}~~~u,v
\ge 0~~~~{\rm Region~ I} \cr
& u = -r~e^\theta~~~~v = r~e^{-\theta}~~~~{\rm for}~~~u < 0, v > 0
{}~~~~{\rm Region~II} \cr & u = - r~e^\theta~~~~v = - r~e^{-\theta}
{}~~~~{\rm for}~~~u,v < 0 ~~~~ {\rm Region~III} \cr & u = r~e^\theta
{}~~~~v = - r~e^{-\theta}~~~~{\rm for}~~~u > 0, v < 0
{}~~~~ {\rm Region~IV}} \eqn\qfivea$$
Regions I and III are the exterior regions. Region II contains the future
black hole singularity at $uv = a$ while Region IV contains the past
singularity.

Let us now obtain $\tT(r,\theta)$ by starting from the definition \qeleven.
This is done by substituting for the explicit forms of $\eta (x,t)$
satisfying Dirichlet boundary conditions and explicitly evaluating the
transform. One gets
$$ \eqalign{& T(r,\theta) = 2 e^{-i\nu \theta} \int_0^\infty dp~K_{i\nu}(2
\smu p)~K_{i\nu} (-2ipr)~~~~~~~\rega \cr &
T(r,\theta) = 2 e^{-i\nu \theta} \int_0^\infty dp~K_{i\nu}(2
\smu p)~K_{i\nu} (2pr)~~~~~~\regb}\eqn\qfiveseven$$
We will not write down formulae for the other two regions since they are
trivially related to those in I and II.
The $T(r,\theta)$ are thus given in terms of $K$-transforms of the
macroscopic loop operator. Both the above integrals may be evaluated
explicitly [\BATEMAN]. The final result is
$$\eqalign{ &T(r,\theta) = {\pi^2~e^{-{\pi \nu \over 2}} \over
4~cosh~\pi\nu} ~\mu^{{i\nu \over 2}}~e^{-i\nu \theta}~(r)^{-(1 +
i\nu)}~F(\half + i\nu, \half; 1; 1 + {\mu \over r^2})~~~~~\rega \cr &
T(r,\theta) = {\pi^2~e^{-{\pi \nu \over 2}} \over 4~cosh~\pi\nu}
{}~\mu^{{i\nu \over 2}}~e^{-i\nu \theta}~(r)^{-(1 +
i\nu)}~F(\half + i\nu, \half; 1; 1 - {\mu \over
r^2})~~~~~\regb}\eqn\qfiveeight$$
These are in exact agreement with {\bf one} of the solutions of \qtwelve\
in each of the regions
\foot{It has been already noted in [\MART] that the solutions of the
liouville model Wheeler de Witt equation may be transformed into solutions
of the tachyon fluctuations in a blackhole background. We have seen,
however, that the matrix model uniquely picks out {\bf one} of the
solutions of the differential equations.}. The second solution of
\qtwelve\ is of the form
$$\eqalign{\tom '(r) = & e^{i \omega \theta}~(r)^{-(1 +i\omega)}
[\Gamma (\half +i\omega)~ \Gamma (\half)~
F(\half + i\omega,\half;1; 1 + {\mu \over r^2})~ {\rm log}~(1 + {\mu \over
r^2}) + \cr & \sum_{m=1}^{\infty} s_m {\Gamma (\half + m + i\omega)
\Gamma (\half + m) \over (m !)^2}~(1 + {\mu \over r^2})^m]
{}~~~~\rega }\eqn\qsixone$$
where $s_m = \sum_{n=1}^m {1 \over n + i\omega - \half} +
{1 \over n - \half} - {2 \over n}$.
In Region II one has to replace $\mu \rightarrow -\mu$.
This solution with a logarithmic singualrity at the position of the
singularity $uv = - \mu$ is not seen in the matrix model.

It is interesting that the $T(r,\theta)$ obtained by evaluating the
transform yields {\underbar{one}}
of the solutions of the differential equation
satisfied by $T(u,v)$. This is related to the fact that the matrix model
always picks out a {\it specific} combination of dressings of the
continuum theory, as is also manifest in the standard interpretation of
the matrix model as a liouville background. Perhaps more significantly the
solution in the interior region which is picked out is the one which is
regular at the singularity $r^2 = \mu$ in Region II, and not the one which
has a logarithmic singularity \foot{The behavior of the tachyon field near
the singularity has been investigated in [\BLACKA] and [\DVV].}.

To obtain the asymptotic states which satisfies physically interesting
boundary conditions in the exterior region it is necessary to rewrite
the above solutions using standard relations between hypergeometric
functions. For example in Region I
$$\eqalign{T(r,\theta) = &{e^{-i\nu \theta} \over r}
[({\mu \over r^2})^{i \nu \over 2}
A(-\nu) F( \half + i\nu, \half; 1 + i\nu; -{\mu \over r^2}) + \cr &
(-1)^{i\nu} ({\mu \over r^2})^{-i \nu \over 2} A(\nu) F(
\half - i\nu, \half; 1 - i\nu; -{\mu \over r^2})]} \eqn\qfivenine$$
where $A(\nu) \equiv {\Gamma(i\nu) \over \Gamma(\half + i \nu) \Gamma
(\half)}$ and we have omitted an overall $\nu$-dependent constant. The two
terms in \qfivenine\ are in exact agreement with two of the
tachyon vertex operators found in the $SL(2,R)/U(1)$ coset model in [\DVV]
after a change of variables $r \rightarrow {\rm sinh}~(r/2)$. These modes
vanish on the past and future null infinities respectively and represent
left and right moving modes at spatial infinity. The other two modes which
vanish at the past and future horizons ( and represent left and
rightmoving plane waves on the horizon) are given by the two terms of the
following rewriting of the hypergeometric functions in \qfiveeight\
$$\eqalign{T(r,\theta) = &{e^{-i\nu \theta} \over r}
[({\mu \over r^2})^{i \nu \over 2}
A(\nu) F( \half - i\nu, \half; 1 - i\nu; -{r^2 \over \mu}) + \cr &
(-1)^{i\nu} ({\mu \over r^2})^{-i \nu \over 2} A(-\nu) F(
\half + i\nu, \half; 1 + i\nu; -{r^2 \over \mu})]} \eqn\qfiveten$$
Thus the transform of the macroscopic loop operator is a linear
combination of the asymptotic solutions in the black hole geometry.
It may be easily checked that the solutions at $r = 0$ approached from
Region I differ from that approached from Region III by a factor $e^{-\pi
\nu}$. This is a signature of the fact that the black hole is in
equilibrium with a thermal bath at temperature $T_{BH} = {1 \over 6 \pi
{\sqrt {\alpha'}}}$.

So far we have been considering the fluctuation $\tT$. What is the
behaviour of the transform of the classical solution, $T_0 (u,v)$ ? It is
easier to evaluate the derivative $\partial_\mu~T_0$.
Using the form of the classical solution for the collective field it may
be easily checked that
$$ \partial_\mu W_0 (p) \equiv \partial_\mu
\int_{2 \smu}^{\infty} dx~e^(-px)~\phi_0 (x) = 2 K_0 (2 \smu p)\eqn\eone$$
Let us now evaluate $\partial_\mu T_0$ in the region containing the future
singularity, Region II. Using the appropriate $(r, \theta)$ coordinates
one gets
$$\partial_\mu T_0 (r,\theta) = \int_0^\infty dp \int_{-\infty}^{\infty}
dt~ e^{2ipr~\sinh~t}~K_0 (2 \smu p) \eqn\etwo$$
The integral over $t$ can be now performed, yielding another modified
Bessel function $K_0 (2pr)$ and finally the integral over $p$ can be
performed using standard K-transforms [\BATEMAN], yielding finally
$$ \partial_\mu~T_0 (r,\theta) = {\pi \over 2r} F(\half,\half;1; 1 - {\mu
\over r^2})~~~~~~~~{\rm Region II} \eqn\sone$$
with similar expressions in region I. As expected the result is
independent of $\theta$. Furthermore {\bf the background is completely
non-singular at the location of the "singularity" at $r = \smu$}.

We have thus shown that the transform of both the background and the
fluctuations are well behaved and nonsingular at the "singularity" already
at the classical level. The implications of this fact is unclear. In
[\DMWBH] the above correspondence between the matrix model and the black
hole has been discussed in the framework of the bosonization of the
fermionic field theory in terms of the quantum distribution function in
phase space $u(x,p,t)$. Consider the bilocal operator of fermions
defined as
$$W(\alpha,\beta,t)= 1/2 \int_{-\infty}^\infty dx\; e^{i\alpha x}\;
\psi^\dagger(x+\beta/2,t)\; \psi(x-\beta/2,t) \eqn\qsixteen$$
In [\DMWBH] one then considers the transform
$${\bar T}(u,v) = \int_{-\infty}^{+\infty} dt \int_{-\infty}^{+\infty}
d\alpha [e^{i \alpha(ue^t - v e^{-t}} W(\alpha,0,t) +
e^{i \alpha(ue^t + v e^{-t}} W(0,\alpha,t) \eqn\stwo$$
The first term is related to our transform with {\it imaginary values of}
$\alpha$. It has been shown that the background value of ${\bar T}$ defined
above is singular at the black hole singularity at the classical level but
the singularity disappears when the exact (all orders) result is used. In
our transform the singularity is not present in the classical level to
begin with. The singularity of ${\bar T}$ at the classical level may be
traced to the fact that it involves macroscopic loops with imaginary
loop lengths.

The crucial point about the transform we defined is that it is {\bf
linear}. This means that the $W_{\infty}$ symmetry discussed in the
previous section is also present in the black hole background. In fact it
has been indeed agrued that the coset model does have a $W_{\infty}$
symmetry [\EGUCHI].

The above relationship between the matrix model and black holes is rather
intruiging. More insight will surely come from the structure of the
interaction terms. In terms of the transformed fields these will be
nonlocal. A better understanding will hopefully allow us to use the exact
results of the matrix model to resolve vexing issues in quantum aspects of
black holes.

\chapter{ACKNOWLEDGEMENTS}

I would like to thank A. Dhar, A. Jevicki, G. Mandal, A. Sengupta
and S. Wadia for discussions and collaborations and A. Sen for a useful
discussion. I thank the organizers of the Spring School at ICTP for their
invitation to present these lectures and their hospitality.

\refout
\end